\documentclass[prl,twocolumn,twoside,preprintnumbers,superscriptaddress,showpacs,showkeys,nofootinbib]{revtex4}
\usepackage{amsmath,slashed}
\usepackage{graphicx,graphics}
\usepackage{dcolumn}
\usepackage[hyperfootnotes=false]{hyperref}
\usepackage{xspace}

\newcommand{\mpi}{M_{\pi}}
\newcommand{\Order}{\mathcal{O}}

\newcommand{\MeV}{\,\text{MeV}}
\newcommand{\GeV}{\,\text{GeV}}
\newcommand{\mN}{m_N}
\newcommand{\mc}{m_\chi}
\newcommand{\Nf}{N_\text{f}}
\newcommand{\N}{\mathcal{N}}
\newcommand{\F}{\mathcal{F}}
\newcommand{\vv}{\mathbf{v}}
\newcommand{\qq}{\mathbf{q}}
\newcommand{\diff}{\text{d}}
\newcommand{\BR}{\text{BR}}
\newcommand{\beq}{\begin{equation}}
\newcommand{\eeq}{\end{equation}}

\begin{document}

\preprint{INT-PUB-17-031}
\title{Improved limits for Higgs-portal dark matter from LHC searches}

\author{Martin Hoferichter}
\affiliation{Institute for Nuclear Theory, University of Washington, Seattle, WA 98195-1550, USA}
\author{Philipp Klos}
\affiliation{Institut f\"ur Kernphysik, Technische Universit\"at Darmstadt, 64289 Darmstadt, Germany}
\affiliation{ExtreMe Matter Institute EMMI, GSI Helmholtzzentrum f\"ur Schwerionenforschung GmbH, 64291 Darmstadt, Germany}
\author{Javier Men\'endez}
\affiliation{Center for Nuclear Study, The University of Tokyo,
	113-0033 Tokyo, Japan}
\author{Achim Schwenk}
\affiliation{Institut f\"ur Kernphysik, Technische Universit\"at Darmstadt, 64289 Darmstadt, Germany}
\affiliation{ExtreMe Matter Institute EMMI, GSI Helmholtzzentrum f\"ur Schwerionenforschung GmbH, 64291 Darmstadt, Germany}
\affiliation{Max-Planck-Institut f\"ur Kernphysik, Saupfercheckweg 1, 69117 Heidelberg, Germany}

\begin{abstract}
Searches for invisible Higgs decays at the Large Hadron Collider
constrain dark matter Higgs-portal models,
where dark matter interacts with the Standard Model fields via the Higgs boson.
While these searches complement dark matter direct-detection experiments,
a comparison of the two limits depends on the coupling of the Higgs boson
to the nucleons forming the direct-detection nuclear target,
typically parameterized in a single quantity $f_N$.
We evaluate $f_N$ using recent phenomenological and lattice-QCD calculations,
and include for the first time the coupling of the Higgs boson to two nucleons
via pion-exchange currents.
We observe a partial cancellation for Higgs-portal models
that makes the two-nucleon contribution anomalously small.
Our results, summarized as $f_N=0.308(18)$,
show that the uncertainty of the Higgs--nucleon coupling
has been vastly overestimated in the past. The improved limits
highlight that state-of-the-art nuclear physics input
is key to fully exploiting experimental searches.
\end{abstract}

\pacs{Dark matter, WIMPs, chiral Lagrangians}
\keywords{95.35.+d, 14.80.Ly, 12.39.Fe}

\maketitle

\section{Introduction}

After the initial discovery of a new particle by the ATLAS and CMS collaborations at the Large Hadron Collider (LHC) in 2012~\cite{Aad:2012tfa,Chatrchyan:2012xdj},
its mass has been further constrained to sub-GeV accuracy, $m_h=125.09(24)\GeV$~\cite{Aad:2015zhl}, and so far neither studies of its spin and parity~\cite{Khachatryan:2014kca,Aad:2015mxa} nor its branching fractions~\cite{Khachatryan:2016vau} have revealed significant deviations from the Standard Model (SM) prediction for the Higgs boson $h$.
An interesting consequence is that
events with large missing transverse momentum would be expected
if the Higgs boson decayed to long-lived non-SM states that do not leave a signature in the detector, so-called invisible Higgs decays.
The absence of such observations sets limits on the invisible decay channels~\cite{Aad:2014iia,Chatrchyan:2014tja,Aad:2014wza,Aad:2015uga,Aad:2015txa,Aad:2015pla,Khachatryan:2016whc},
which provide stringent constraints on beyond-SM physics.

One particular example concerns Higgs-portal models for dark matter, see, e.g.,~\cite{Kanemura:2010sh,Djouadi:2011aa,Djouadi:2012zc,Beniwal:2015sdl}, in which the dark-matter candidate $\chi$, a weakly interacting massive particle (WIMP), interacts with the SM fields via exchange of the Higgs boson.
With the nature of dark matter still a major puzzle, the quest to understand its
composition is being vigorously pursued, besides collider signatures, in direct-detection experiments
looking for the WIMP scattering off atomic nuclei,
as well as indirect searches~\cite{Baudis:2016qwx}.
As long as the WIMP mass fulfills $\mc < m_h/2$, limits on the invisible decay width of the Higgs boson are in a one-to-one correspondence with direct-detection limits
on the WIMP--nucleon cross section $\sigma_{\chi N}$, provided the nuclear physics input is sufficiently under control.

\begin{figure}[t]
 \centering
 \includegraphics[height=2.1cm,clip]{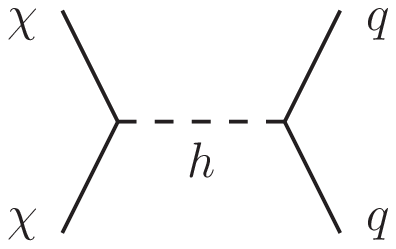}\qquad
 \includegraphics[height=2.1cm,clip]{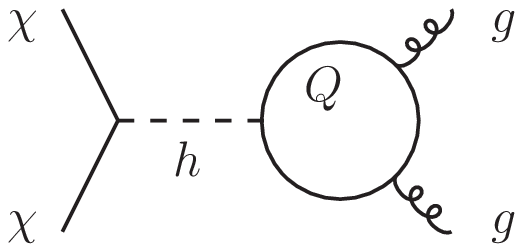}
 \caption{Higgs-mediated interaction of the WIMP $\chi$ with the light quarks $q=u,d,s$ (left) and the gluon field $g$ by closing a heavy-quark loop $Q=c,b,t$ (right).}
 \label{fig:qhgh}
\end{figure}

Such comparison relies on the
coupling of the Higgs boson to a single nucleon,
which proceeds via the nucleon matrix elements for the scalar current 
\beq
\label{scalar_coupling}
\mN f_q^N=\langle N|m_q\bar q q|N\rangle
\eeq
for the light quarks $q=u,d,s$ (with quark masses $m_q$ and nucleon mass $\mN$), as well as through heavy-quark loops coupling to gluon fields, see Fig.~\ref{fig:qhgh}. Integrating out the heavy quarks at leading order in the strong coupling constant $\alpha_s$ leads to an effective coupling~\cite{Shifman:1978zn}
\beq
\label{f_N_leading_order}
f_N=\sum_{q=u,d,s,c,b,t}f_q^N=\frac{2}{9}+\frac{7}{9}\sum_{q=u,d,s}f_q^N
\eeq
to describe the Higgs--nucleon interaction. Both recent LHC analyses constraining Higgs-portal dark matter from invisible Higgs decays~\cite{Aad:2015pla,Khachatryan:2016whc} use a central value $f_N=0.326$~\cite{Young:2009zb} with variations covering $f_N=0.260\ldots 0.629$ (motivated by~\cite{Toussaint:2009pz}, see~\cite{Djouadi:2011aa}; see also~\cite{Sage:2015wfa}).
Given recent progress in lattice-QCD calculations~\cite{Durr:2015dna,Yang:2015uis,Abdel-Rehim:2016won,Bali:2016lvx} and pion--nucleon ($\pi N$) phenomenology~\cite{Alarcon:2011zs,Hoferichter:2015dsa,Hoferichter:2016ocj,RuizdeElvira:2017stg}, this large range in $f_N$ no longer reflects
the current knowledge of the scalar couplings of the nucleon.
Consequently, the limits for $\sigma_{\chi N}$ derived from the searches for invisible Higgs decays can be significantly improved. In the first part of this paper, we provide a more detailed assessment of the current situation, including corrections to Eq.~\eqref{f_N_leading_order} regarding isospin violation and higher orders in $\alpha_s$ for the heavy-quark loops.

In addition, the standard decomposition~\eqref{f_N_leading_order}
does not reflect that atomic nuclei
are strongly interacting many-nucleon systems,
so that not only the one-body (single-nucleon) matrix elements enter.
As first pointed out in~\cite{Prezeau:2003sv}, corrections to this picture from two-body currents, where the Higgs couples to a pion exchanged between two nucleons, can become important. Such corrections are conveniently addressed within chiral effective field theory (EFT)~\cite{Epelbaum:2008ga,Machleidt:2011zz}, an expansion around the chiral limit of QCD in terms of momenta and quark masses, as successfully demonstrated in
previous applications to WIMP--nucleus scattering~\cite{Cirigliano:2012pq,Menendez:2012tm,Klos:2013rwa,Baudis:2013bba,Cirigliano:2013zta,Hoferichter:2015ipa,Hoferichter:2016nvd,Korber:2017ery} 
(related work using chiral EFT for WIMP--nucleon interactions~\cite{Hoferichter:2015ipa}
is restricted to the coupling to a single nucleon~\cite{Bishara:2016hek,Bishara:2017pfq} or focuses on the
calculation of the nuclear states based on chiral EFT forces~\cite{Gazda:2016mrp}).
In this work, 
we extend previous calculations limited to xenon~\cite{Hoferichter:2016nvd} to a broad range of target nuclei and show that the corrections from the coupling to two nucleons are largely coherent and can 
thus be absorbed into a redefinition of the single-nucleon $f_N$.
The combination of the one- and two-body contributions to $f_N$ then
allows us to compare the limits on Higgs-portal models
from collider and direct-detection experiments using state-of-the-art
nuclear physics input.

\section{Scalar couplings of the nucleon}

The evaluation of the quark-level operators from Fig.~\ref{fig:qhgh} between nucleon states requires knowledge of the scalar couplings defined in Eq.~\eqref{scalar_coupling}. 
To arrive at Eq.~\eqref{f_N_leading_order}, the couplings $f_Q$ of the heavy quarks $Q=c,b,t$ are eliminated in favor of the light quarks' according to the following procedure~\cite{Shifman:1978zn}:
at $\Order(\alpha_s)$, the QCD trace anomaly for $\Nf$ active degrees of freedom reads
\beq
\theta^\mu_\mu=\sum_{i=1}^{\Nf}m_{q_i} \bar q_i q_i-\bigg(11-\frac{2\Nf}{3}\bigg)\frac{\alpha_s}{8\pi}G_{\mu\nu}^aG^{\mu\nu}_a, 
\eeq
with gluon field-strength tensor $G_{\mu\nu}^a$. Integrating out the heavy quarks then yields for each flavor
\begin{align}
\mN f_Q^N&=\langle N|m_Q\bar Q Q|N\rangle =-\frac{\alpha_s}{12\pi} \langle N|G_{\mu\nu}^aG^{\mu\nu}_a|N\rangle\notag\\
&=\frac{2}{27}\bigg(\langle N|\theta^\mu_\mu|N\rangle-\sum_{q=u,d,s}\langle N|m_q\bar q q|N\rangle\bigg),
\end{align}
so that 
\beq
\label{fQ}
f_Q^N=\frac{2}{27}\bigg(1-\sum_{q=u,d,s} f_q^N\bigg)+\Order(\alpha_s)
\eeq
and Eq.~\eqref{f_N_leading_order} follows.  

Modifications of the leading result~\eqref{f_N_leading_order} arise from two sources: isospin-breaking and perturbative corrections. For $u$- and $d$-quarks the dominant isospin-breaking effects cancel in the sum $f_u^N+f_d^N$, with remaining effects of the size~\cite{Crivellin:2013ipa} 
\beq
f_p-f_n=\frac{(m_p-m_n)^\text{str}}{\mN},
\eeq
where $(m_p-m_n)^\text{str}$ denotes the portion of the proton--neutron mass difference proportional to $m_d-m_u$. In a nucleus, this would lead to a relative correction
\beq
\label{IV}
\frac{\Delta f_N^\text{IV}}{f_N}=\frac{Z-N}{A}\frac{f_p-f_n}{2f_N}\lesssim 0.1\%,
\eeq
where $Z$/$N$ refer to proton/neutron number, $A=Z+N$, and we have inserted values for a typical Xe isotope. Thus, isospin-violating effects in $f_N$ are small~\cite{Crivellin:2015bva} and will be neglected in the following.
The perturbative corrections have been worked out in detail in~\cite{Hill:2014yxa}. Here, we use the full result for the $c$ quark
\beq
f_c^N=0.083-0.103\sum_{q=u,d,s}f_q^N+\Order\big(\alpha_s^4,m_c^{-1}\big),
\eeq
as well as the $\Order(\alpha_s)$ correction for $q=b,t$. 

\begin{table*}[t]
\renewcommand{\arraystretch}{1.3}
\centering
\begin{tabular}{crrrrrrrrr}\toprule
Ref. & $f_u^N+f_d^N$ & $f_s^N$ & $f_c^N$ & $f_Q^N$ & $f_N$ & $\hat f_c^N$ & $\hat f_b^N$ & $\hat f_t^N$ & $\hat f_N$\\\colrule
\cite{Durr:2015dna} & $41(5)$ & $113(60)$ &  & $63(4)$ & $342(47)$ & $67(6)$ & $65(5)$ & $64(5)$ & $349(44)$\\
\cite{Yang:2015uis} & $49(8)$ & $43(13)$ &  & $67(1)$ & $294(12)$ & $74(2)$ & $70(1)$ & $68(1)$ & $304(11)$\\
\cite{Abdel-Rehim:2016won} & $40(5)$ & $44(11)$ & $85(25)$ & $68(1)$ & $287(10)$ & $74(1)$ & $71(1)$ & $69(1)$ & $298(9)$\\
\cite{Bali:2016lvx} & $37(6)$ & $37(13)$ &  & $69(1)$ & $280(11)$ & $75(1)$ & $72(1)$ & $70(1)$ & $291(11)$\\\colrule
\cite{Hoferichter:2015dsa} & $63(4)$ &  &  & &\\
\botrule
\end{tabular}
\caption{Lattice results~\cite{Durr:2015dna,Yang:2015uis,Abdel-Rehim:2016won,Bali:2016lvx} for the scalar couplings of the nucleon (in units of $10^{-3}$), for the lightest quarks compared to the result from $\pi N$ scattering~\cite{Hoferichter:2015dsa}. $f_Q^N$ and $f_N$ refer to Eqs.~\eqref{fQ} and~\eqref{f_N_leading_order}, respectively, while quantities with a hat include higher-order perturbative corrections as described in the main text. Statistical and systematic lattice errors have been added in quadrature.}
\label{tab:lattice}
\end{table*}

In this way, we are left with the determination of the scalar couplings of the light quarks in the isospin limit. In general, these couplings cannot be measured in experiment due to the absence of a scalar source, leaving lattice QCD as the only viable option. However, an important exception concerns the scalar couplings of $u$- and $d$-quarks, which the Cheng--Dashen low-energy theorem~\cite{Cheng:1970mx,Brown:1971pn} relates to the $\pi N$ scattering amplitude, offering a unique opportunity to test the results of lattice-QCD calculations with experiment.
More precisely, the low-energy theorem allows one to extract the $\pi N$ $\sigma$-term $\sigma_{\pi N}=\mN(f_u^N+f_d^N)$ from the isoscalar amplitude evaluated at a particular kinematic point in the subthreshold region. The required analytic continuation can be performed in a stable manner using constraints from analyticity and unitarity in the form of Roy--Steiner equations~\cite{Hoferichter:2015dsa,Ditsche:2012fv,Hoferichter:2012wf,Hoferichter:2015tha,Hoferichter:2015hva,Hoferichter:2016duk,Siemens:2016jwj},
with the result that ultimately the value of $\sigma_{\pi N}$ is fully determined by the amplitude at threshold, i.e.\ the $\pi N$ scattering lengths. These, in turn, are known to high accuracy from measurements in pionic atoms~\cite{Gotta:2008zza,Strauch:2010vu,Hennebach:2014lsa,Baru:2010xn,Baru:2011bw}. The result $\sigma_{\pi N}=59.1(3.5)\MeV$~\cite{Hoferichter:2015dsa} agrees with calculations based on chiral perturbation theory, once the low-energy constants are extracted from a phase-shift solution consistent with the pionic-atom scattering lengths~\cite{Alarcon:2011zs}.

Unfortunately, these results from $\pi N$ phenomenology are in tension with the most advanced lattice calculations~\cite{Durr:2015dna,Yang:2015uis,Abdel-Rehim:2016won,Bali:2016lvx}, see Table~\ref{tab:lattice}.
These results correspond to (nearly) physical quark masses, thereby eliminating a major source of systematic uncertainty in earlier calculations, but finite-volume corrections, discretization effects, and
excited-state contamination may not be under sufficient control yet.
It has also been suggested that the tension could be due to deficient input for the $\pi N$ scattering lengths in the phenomenological analysis, which could again be checked on the lattice~\cite{Hoferichter:2016ocj}, but the fact that low-energy scattering data corroborate the pionic-atom values makes this explanation appear unlikely~\cite{RuizdeElvira:2017stg}.

For $f_s^N$ a similar cross check of lattice results with phenomenology would need to rely on $SU(3)$ relations, but the slow convergence of the $SU(3)$ expansion prohibits a meaningful, quantitative test. Table~\ref{tab:lattice} also lists the heavy-quark couplings as well as the resulting values for $f_N$ with and without perturbative corrections, which prove to be only of moderate size. In particular, the direct determination of $f_c^N$ from~\cite{Abdel-Rehim:2016won} is consistent with the perturbative calculation.

To combine the results of Table~\ref{tab:lattice} into a single value for $f_N$, we perform a naive average of the four lattice values, take $f_u^N+f_d^N=0.052(12)$ as the mean between lattice and phenomenology with an error sufficiently large to cover both, and increase the error in $f_s^N=0.043(20)$ to be able to accommodate a similar potential bias as for $f_u^N+f_d^N$.
We include the perturbative corrections, but add an additional error $\Delta f_N^\text{pert}=0.005$, about half the shift observed in Table~\ref{tab:lattice} when adding the higher-order terms. Altogether, this leads to the one-body (1b) contribution
\beq
\label{fN1b}
f_N^\text{1b}=0.307(9)_{ud}(15)_{s}(5)_\text{pert}=0.307(18).
\eeq

\section{Two-body currents}

\begin{figure}[t]
 \includegraphics[width=\linewidth,clip]{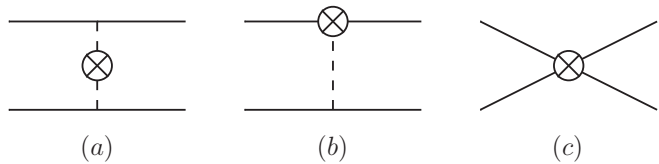}
 \caption{Two-body contributions to WIMP--nucleus scattering. Solid/dashed lines refer to nucleons/pions, the crosses to the coupling of the external current.}
 \label{fig:2b}
\end{figure}

The spin-independent limits on $\sigma_{\chi N}$ derived from direct-detection experiments, see, e.g., \cite{Angloher:2015ewa,Agnese:2015nto,Agnes:2015ftt,Amole:2016pye,Armengaud:2016cvl,Tan:2016zwf,Akerib:2016vxi,Aprile:2017iyp},
assume a simple relation with the differential WIMP--nucleus cross section 
\beq
\label{diff_cs}
\frac{\diff \sigma_{\chi\N}}{\diff\qq^2}=\frac{\sigma_{\chi N}}{4\vv^2\mu_N^2}\F^2(\qq^2),\qquad \mu_N=\frac{\mN\mc}{\mN+\mc},
\eeq
where $\qq$ is the momentum transfer, $\vv$ the relative velocity, and $\F(\qq^2)$ a nuclear form factor normalized to $\F(0)=A$.
This decomposition ignores subleading corrections, such as isospin violation or two-body effects, which in general all come with an independent nuclear form factor~\cite{Hoferichter:2016nvd}.
For the special case of Higgs-mediated WIMP--nucleus scattering, isospin-breaking corrections are small, see Eq.~\eqref{IV}.
In contrast, the leading effect of two-body currents is expected to be coherent
(scaling with $A$), so that it can be included as an effective shift in $f_N$.
This contribution is crucial to ensure consistency:
direct-detection limits on $\sigma_{\chi N}$ convert limits
on the WIMP--nucleus rate to an effective WIMP--nucleon cross section based on Eq.~\eqref{diff_cs},
effectively subsuming the coupling to two nucleons.
The cross section derived from limits on invisible Higgs decays needs to be consistent with this convention, and thus
has to include the two-nucleon effects as well. 
Note that if the nuclear structure factors used in the interpretation of the direct-detection experiments 
accounted for two-body contributions (as is the case for the recent spin-dependent limits~\cite{Aprile:2013doa,Uchida:2014cnn,Fu:2016ega,Akerib:2017kat}),
the resulting WIMP--nucleon cross section could be interpreted as a true
one-body quantity, and then the transition from collider limits would not require adding a two-body correction.

\begin{figure*}[t]
 \includegraphics[width=12cm,clip]{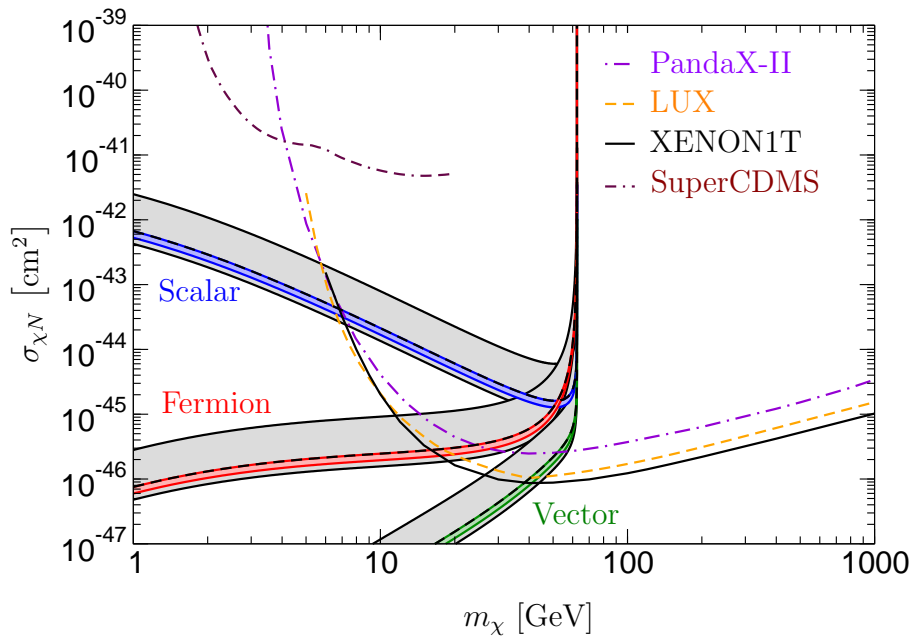}
 \caption{Exclusion limits for scalar (blue), fermion (red), and vector (green) Higgs-portal WIMPs. The gray-shaded bands refer to the range $f_N=0.260\ldots 0.629$ from the most recent ATLAS~\cite{Aad:2015pla} and CMS~\cite{Khachatryan:2016whc} analyses, the dashed lines to the central value $f_N=0.326$ considered therein, and the colored bands to our improved limits using Eq.~\eqref{fN_final}. For comparison, we show the direct-detection limits from  SuperCDMS~\cite{Agnese:2015nto}, PandaX-II~\cite{Tan:2016zwf}, LUX~\cite{Akerib:2016vxi}, and XENON1T~\cite{Aprile:2017iyp}.}
 \label{fig:LHC}
\end{figure*}

The leading two-body diagrams in chiral EFT are shown in Fig.~\ref{fig:2b}. 
In the derivation, the couplings to the scalar current $m_q\bar q q$ and the trace anomaly $\theta^\mu_\mu$  lead to two different responses~\cite{Hoferichter:2016nvd}. For the scalar case, diagram $(b)$ is suppressed
by two orders because there is no scalar source in the leading $\pi N$ chiral Lagrangian, while the subleading Lagrangian does not produce a single-pion vertex.
The validity of the underlying chiral EFT counting could be checked by comparing to nuclear $\sigma$-terms 
from lattice QCD, see~\cite{Beane:2013kca} for such a calculation at heavy pion masses.
Similarly, diagram $(c)$ requires an insertion of the quark-mass matrix, which also suppresses
this contribution by two orders in the chiral counting.
This leaves only the coupling to the pion in flight in diagram $(a)$.
For $\theta^\mu_\mu$, diagram $(a)$ also enters, diagram $(b)$ is again only next-to-next-to-leading order, but there is a leading-order contribution to diagram $(c)$, which was not included in~\cite{Hoferichter:2016nvd}. 

The terms in diagram $(c)$ are directly related to the $(N^\dagger N)^2$ contact operators that
enter in the $NN$ potential. Indeed, one can show that combining them with parts of diagram
$(a)$, they yield the leading $NN$ potential $V_{NN}$~\cite{Hoferichter:2017}. Taken together with momentum-dependent one-body corrections at the same order, which can be identified with the kinetic-energy operator $T$, one finds a combined contribution for the $\theta^\mu_\mu$ response proportional to
\beq
\label{SE}
\langle \Psi|T+V_{NN}|\Psi\rangle=E_\text{b},
\eeq
where $E_\text{b}<0$ is the binding energy of the nucleus
represented by $|\Psi\rangle$, obtained from $NN$ interactions only. This adds to the remaining parts from diagram $(a)$ (both the scalar and $\theta^\mu_\mu$ responses) to an effective two-body shift in $f_N$~\cite{Hoferichter:2016nvd,Hoferichter:2017}
\beq
\label{fN2b}
f_N^\text{2b}=\frac{1}{A}\frac{\mpi}{\mN}\frac{11}{9}\F_\pi(0)-\frac{4}{9}\frac{E_\text{b}}{A\mN},
\eeq
where $\F_\pi(\qq^2)$ refers to the nuclear structure factor for the scalar current~\cite{Hoferichter:2016nvd}, in the same normalization where $\F(0)=A$ in Eq.~\eqref{diff_cs}.

We have evaluated the two-body contribution $\F_\pi(0)$
in harmonic-oscillator states using the occupation numbers that result
from a large-scale shell-model diagonalization
with state-of-the-art nuclear interactions
(see~\cite{Hoferichter:2016nvd,Hoferichter:2017} for details)
for a wide range of nuclei
that includes all stable xenon, argon, germanium, and silicon isotopes.
We observe a robust coherence proportional to $A$
of the two-nucleon structure factors, leading to
\begin{align}
\label{2b_final}
f_N^\text{2b}&=\big[-3.2(0.2)(2.1) + 5.0(0.4)\big]\times 10^{-3}\notag\\
&=1.8(2.1)\times 10^{-3},
\end{align}
where the two terms correspond to those in Eq.~\eqref{fN2b}, and in each term the first uncertainty is from the variation over the different isotopes. For the binding-energy term, we take the experimental energy, 
corrected for
the Coulomb contribution following~\cite{Duflo:1995ep}. This effectively includes higher-order terms, e.g., due to $3N$ interactions. For the first term in Eq.~\eqref{fN2b}, the second uncertainty is due to the truncation in the chiral expansion and from the use of the harmonic-oscillator model to evaluate $\F_\pi(0)$. We have taken this uncertainty significantly larger than that naively estimated by the chiral expansion to account for possible cancellations of different contributions, as would be manifest between kinetic and potential energies.
We have also estimated the uncertainty in the many-body calculation
by using the occupation numbers corresponding to two nuclear interactions,
and the effect is less than $1\%$. 

Numerically, we observe a partial cancellation between the two terms, to the effect that the size of the two-body corrections
is reduced from the expected few-percent level to below $1\%$.
Combined with the Higgs coupling to a single nucleon, we find
\beq
\label{fN_final}
f_N=f_N^\text{1b}+f_N^\text{2b}=0.308(18),
\eeq
which is our final result.

\section{Impact on LHC exclusion limits}

The relation between limits for the branching fraction of invisible Higgs decays $\BR_\text{inv}$ and $\sigma_{\chi N}$ depends on the nature of the Higgs portal. For instance, 
for a scalar, vector, or fermion
portal $H^\dagger H\, S^2$, $H^\dagger H\, V_\mu V^\mu$, $H^\dagger H\, \bar{f}f$, respectively,
with Higgs doublet $H$ and scalar, vector, or fermion fields $S$, $V^{\mu}$, $f$, one finds
\beq
\sigma_{\chi N}=\Gamma_\text{inv}\frac{8\mN^4f_N^2}
{v^2\beta m_h^3(\mc+\mN)^2}g_\chi\bigg(\frac{m_h}{\mc}\bigg),
\eeq
where $g_S(x)=1$, $g_V(x)=4/(12-4x^2+x^4)$, and $g_f(x)=2/(x^2-4)$,
$\beta=\sqrt{1-4\mc^2/m_h^2}$, $v=246\GeV$ is the Higgs vacuum expectation value, and $\BR_\text{inv}=\Gamma_\text{inv}/(\Gamma_\text{inv}+\Gamma_\text{SM})$, with $\Gamma_\text{SM}=4.07\MeV$~\cite{Heinemeyer:2013tqa}.
For details
we refer to~\cite{Djouadi:2011aa}.
 
The consequences of the improved coupling $f_N$ are illustrated in Fig.~\ref{fig:LHC}, which is inspired by analogous figures in the most recent ATLAS~\cite{Aad:2015pla} and CMS~\cite{Khachatryan:2016whc} studies. For definiteness, we adopt the CMS limit $\BR_\text{inv}<0.20$ (at $90\%$ confidence level).
Even on a logarithmic scale the impact
due to the improved nuclear physics input used is striking.
Only a narrow band remains in the previously considered parameter space, tightening considerably the limits that result from experimental searches.
Our results highlight that a careful determination
of the nuclear physics related to the dark matter interactions
is key for a correct determination of
the limits that searches for invisible Higgs decays impose on Higgs-portal models for dark matter.

In summary, we have evaluated the Higgs--nucleon coupling, $f_N$,
using state-of-the-art phenomenological and lattice-QCD  calculations,
and including the coupling of the Higgs to two nucleons
as predicted by chiral EFT.
Our result, $f_N=0.308(18)$, reduces dramatically
the uncertainty in the excluded region derived from experiment.
This highlights that the nuclear physics input
is key to fully exploiting the consequences of experimental searches, and
that it needs to be treated consistently in limits from collider
experiments and direct detection.

\section*{Acknowledgments}
\begin{acknowledgments}
We thank Hans-Werner Hammer, David B.~Kaplan, and Jordy de Vries for valuable discussions.
Financial support by  the DOE (Grant No.\ DE-FG02-00ER41132), the DFG (Grant SFB 1245), the
ERC (Grant No.\ 307986 STRONGINT), MEXT (Priority Issue on Post-K computer), and JICFuS is gratefully acknowledged.
AS thanks the INT for its hospitality during the completion of this work.
\end{acknowledgments}

\end{document}